# What is the Q of a Blackbody?
A small contribution to Gustav Robert Kirchhoff's bicentennial (1824 – 1887)

Arthur Ballato[1,a] and John Ballato[2,b]

[1]Holcombe Department of Electrical and Computer Engineering, Clemson University, Clemson, SC 29634
[2]Department of Materials Science and Engineering, Clemson University, Clemson, SC 29634
Author Emails: jballat@clemson.edu, art.ballato@gmail.com

**Abstract**
The blackbody spectrum "half-power points" are used to assign effective Q "quality factor" values that are found to be less than unity whether frequency or wavelength scaling is used. A comparison with values for coherent oscillators is made. This exercise blends two of Kirchhoff's interests, and is instructive in its own right, as it bridges the often mutually exclusive engineering and scientific disciplines.

**I. INTRODUCTION**
In 1860, the mathematical physicist Gustav Kirchhoff coined the term "blackbody" to denote an ideal surface that absorbs all incident electromagnetic radiation.[1] Fifteen years earlier, as a student, he had introduced the two eponymous circuit laws now universally familiar to electrical engineering (EE) students.[2][F1,F2] This paper establishes another link between these disciplines with which he was so conversant, and celebrates the bicentennial of his birth. Connecting blackbody physics and circuit engineering is apposite, not only for didactic reasons, but because it serves also as a reminder that not a few notable scientists began their careers with an engineering education, e.g., Röntgen, Debye, Dirac, Onsager, Bardeen, etc.

Studies of blackbody (BB) radiation led, as is very well known, to quantum mechanics, and so much has been written about it at all levels that it might be asked if anything new could, (or should!), be added.[3-35] We answer in the affirmative. It is both pedagogically interesting and apt to apply an EE concept to quantify BB radiation. Prescinding from the legitimate objection that such a disparate application, whereby the Q concept usually applied to a coherent resonance should be applied to a completely incoherent photon fluid, we nevertheless find the result to be fully consonant with the exceedingly low values intuitively expected, but unexpectedly to be independent of the blackbody temperature.[F3]

We begin by discussing briefly lumped electrical circuits, and introduce the concept of Q from an EE circuit point of view. This is followed by a bare-bones sketch of BB. Finally, the equivalent Qs of BB spectra are evaluated.

**II. LUMPED ELECTRICAL CIRCUITS – EE 101**
It can be said that with the introduction of electrostatic generators to produce electric charge, and Leiden jars for its storage, the science of electricity began its steps to maturity. Volta's batteries [36] subsequently permitted production of steady electrical currents, and led to the laws of Ørsted,[37] Ampère,[38]-[40], Ohm,[41] and, of course, Kirchhoff. These lumped circuit developments preceded the magisterial unification of electrodynamics at the hands of Faraday and Maxwell with the introduction of field concepts.[42]-[46] Today, the historical roles are reversed, and lumped circuits are considered to follow logically from the more general concept of fields.[47] The correspondences between the Maxwellian (field) and the EE (lumped) approaches have been discussed by Hansen[48], Dicke[49], and particularly by Fano, Chu, and Adler, [50] as well as by others.[51]-[54] Lumped circuit elements (capacitors, inductors, and resistors),[55] represented by graphical symbols, and their mutual attachments to form circuits, follow from the assumption of unbounded lightspeed ($1/c \to 0$), and as such they have no spatial extensions, nor do they mirror the physical geometry of the system represented. They are idealized repositories of electric and magnetic energy and an element of dissipation, with associated equations: $i = C \cdot (de/dt)$ (capacitor), $e = L \cdot (di/dt)$ (inductor), and $e = R \cdot i$ (resistor). In these equations, i is the current through the element, e is the voltage across the element, C the capacitance, L the inductance, R the resistance, and d/dt the time operator.

Nature proceeds in the time domain, and its descriptions take the form of differential equations (DEs). Not infrequently it is more readily interpreted by transformations into the frequency domain, whereby DEs are converted to algebraic functions. In the case of electrical networks, the DEs characterize generalized Ohm's laws relating voltage (e) across, and current (i) through an element, and the Laplace transform is used to convert these DEs of individual circuit elements into algebraic functions relating the independent and dependent variables, i.e., the input-output relations.[F4] The forerunner of this method, "operational calculus," was largely developed by Heaviside.[56]-[63]

These individual circuit elements are then combined, in accordance with the topology of the given network, to produce network input-output relations that take the form of rational functions, i.e., quotients of finite polynomials – a form more easily manipulated and interpreted than the original DEs.[F5] The boundary conditions on the DEs are simply Kirchhoff's laws imposing constraints on currents at each junction of elements (nodes), and on voltages around each complete loop of elements. Circuit theory is now a mature field, having come a long way from its infancy with Kirchhoff.[64-70]

### III. THE CONCEPT OF Q AS A QUALITY FACTOR
The lumped circuit elements L and C store magnetic and electric energy, respectively, while R, representing loss, dissipates energy. In circuit configurations comprised of R, L, and C elements, it is found that the response to steady-state excitation varies with frequency. At particular frequencies, corresponding to solutions of the homogeneous DEs characterizing the configuration (complementary DE solutions), the responses reach local extrema, limited only by the presence of loss. Quantifying these "resonances" in magnitude and frequency extent led to the concept of circuit Qs. Johnson [71] first used the symbol "Q", while Legg and Given [72] coined the term "quality factor." Green [73] gives an excellent account of its early history. A search for an alternative etymology of "Q" did not succeed.[74]-[85]

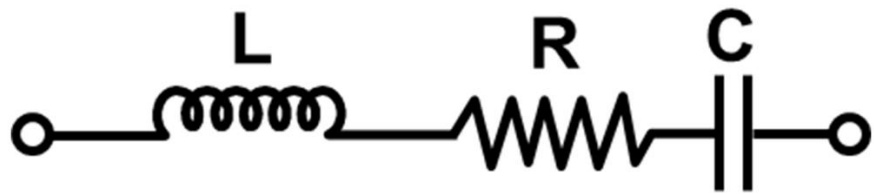

Fig. 1. (a) The series RLC circuit. Resistor R is the dissipative component, evolving power (heat). Inductor L and capacitor C are lossless components that store energy.

Equivalent definitions of Q as a selectivity parameter quantifying the sharpness of resonances appear in the EE literature.[6],[48],[50],[86],[87] Some follow from various energy or power relations, e.g. $Q = \omega \cdot$

$(\langle E_{electric} + E_{magnetic}\rangle/\langle P_{dissipated}\rangle)$; $Q = \omega \cdot$ (Peak stored energy/Average power); $Q \propto$ (Energy stored/Energy dissipated per cycle); or as frequency derivatives of immittance functions: $Q \propto \omega \cdot \partial(\text{susceptance})/\partial\omega \propto \omega \cdot \partial(\text{reactance})/\partial\omega$; or as the shape factor of a resonance curve: $Q = f_o/\Delta f$; or simply as a circuit relation: $Q = (1/R) \cdot \sqrt{(L/C)}$.[48],[88] Alternatively, it may be defined as $Q = \pi/\delta$, where $\delta$ is the logarithmic decrement when the circuit is subjected to a transient excitation in the time domain. Additional aspects are discussed by Feld [89], and particularly by Ohira [90],[91], and references therein. The concept of Q is also relevant in connection with probability distribution functions; see Table VI.

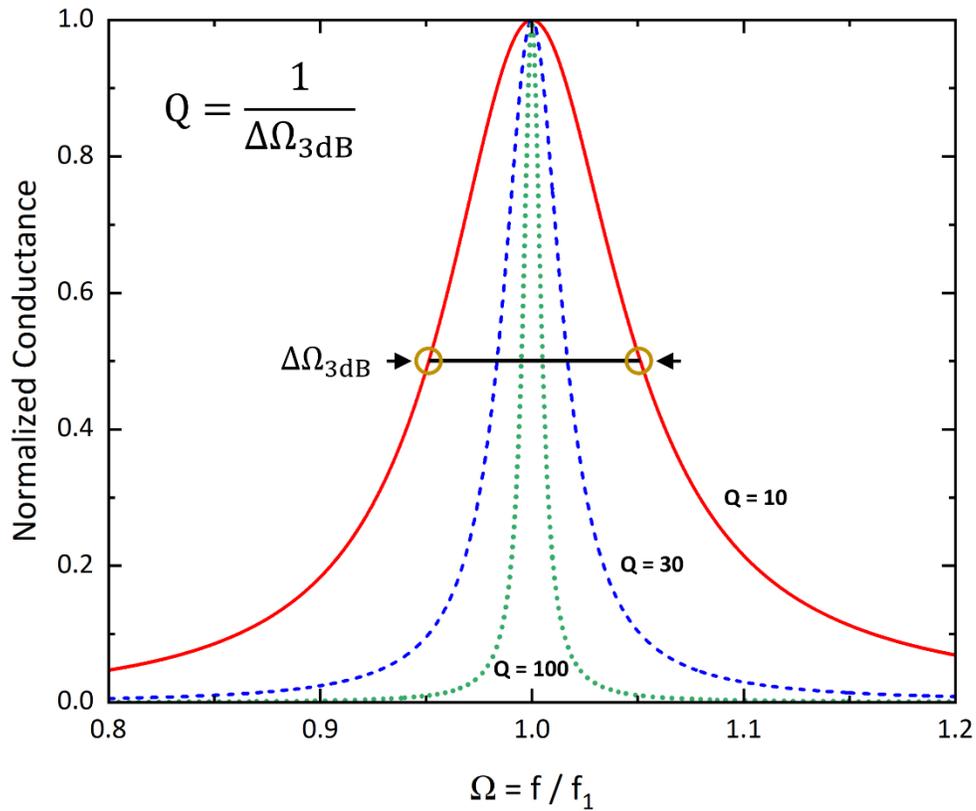

Fig. 1. (b) Resonance curve of the series RLC circuit, with 3dB points shown. Frequency is defined as $f = \omega/2\pi$. Resonance frequency $f_1$ equals $\omega_1/2\pi$, where $L \cdot C \cdot \omega_1^2 \equiv 1$, and normalized frequency is $\Omega = \omega \cdot \sqrt{(L \cdot C)} = f/f_1$. Quality factor, $Q = (1/R) \cdot \sqrt{(L/C)} = 1/\Delta\Omega_{3dB}$. When discussing BB, the EE symbol "f" is replaced by "ν." As a sop to the EEs, shown are the "3dB points," rather than the true ½-power points. These are related by $10 \cdot \log_{10}(2) \approx 3.0103$ dB.

**A. Application to the series RLC circuit**

In the case where lumped elements R, L, and C are in series as in Fig. 1. (a), an impressed voltage e, with $\exp(j\omega t)$ temporal variation, placed across the end terminals, will produce a steady-state common current i having the same temporal variation, albeit generally with a different phase with respect to that of the voltage. The complex input impedance is defined to be $Z = e/i$; complex admittance is defined as $Y = 1/Z$ (the term "immittance" is used as a general term for either Z or Y). While the R, L, and C values are constants, the immittance is a function of frequency, as also, in general, are its real and imaginary parts. These are defined as: resistance = $\text{Re}(Z)$, reactance = $\text{Im}(Z)$, conductance = $\text{Re}(Y)$, and

susceptance = $\text{Im}(Y)$. With the further definitions of normalized frequency $\Omega$, and quality factor Q as: $\Omega = \omega \cdot \sqrt{(L \cdot C)}$ and $Q = (1/R) \cdot \sqrt{(L/C)}$, the Laplace-transformed expressions for input impedance, normalized to R, is $Z/R = [1 + j \cdot Q \cdot (\Omega - 1/\Omega)]$.[92] It then follows that normalized conductance, $\text{Re}(R \cdot Y) = g = 1/[1 + Q^2 \cdot (\Omega - 1/\Omega)^2]$. This is plotted in Fig. 1. (b) as function of $\Omega$ for a number of Q values.[92] The dissipated power is proportional to g, and is a maximum at $\Omega$ = 1, the resonance frequency; this is the eigenvalue of the corresponding DE for the circuit, now appearing as the root of a polynomial.

**B. The half-power points**
The frequency expanse, centered about the resonance peak, that is consonant with the definitions of Q given above, is shown in Fig. 1. (b) as $\Delta\Omega_{3dB}$. This is the 3-dB "bandwidth" found in the EE literature. More accurately, the bandwidth is reckoned as the frequency difference between the two points where the dissipated power in the circuit is one-half of the maximum.[93] The half-power (physics) and 3-dB (EE) points are related by $10 \cdot \log_{10}(2^{\pm 1}) \approx \pm 3.0103 \text{dB}$. For the series (or parallel) RLC circuit, the half-power points are determined from the condition g = ½, yielding two frequencies $\Omega^{(\pm)} = \sqrt{(1 + 1/(2 \cdot Q)^2)} \pm 1/2Q$. These are related by the usual definition: $Q = 1/[\Omega^{(+)} - \Omega^{(-)}] = 1/\Delta\Omega = (1/R) \cdot \sqrt{L/C}$.

Table I. Table of linewidths and Q values for various resonant structures. The Hg linewidth assumes radiation damping only. Other tables are given in Green[73] and in Smith.[77]

| System | Q | Remarks | Reference |
|---|---|---|---|
| Golf ball | 10 | $C_R$ = 85% [94] | [73],[77] |
| GaN LED | 25 | 16nm linewidth @ 400nm | [95],[96] |
| BK7 glass | 700 | 0.3 – 2.5 µm band | [93] |
| Nd:YAG glass laser | $1.3 \cdot 10^{+3}$ | 210 GHz linewidth @1064nm | [96] |
| Ruby laser | $7.2 \cdot 10^{+3}$ | 60 GHz linewidth @ 694.3nm | [96] |
| He-Ne gas laser | $3.2 \cdot 10^{+5}$ | 1.5 GHz linewidth @ 632.8nm | [96] |
| Si wineglass resonator | $4 \cdot 10^{+6}$ | 2 MHz @ RT | [101],[102] |
| Quartz resonators | $5 \cdot 10^{+6}$ | 2.5 – 5 MHz @ RT | [97]-[100] |
| Quartz resonators | $50 \cdot 10^{+6}$ | 2.5 – 5 MHz @ 4.1K | [97]-[100] |
| Green Hg line | $4.6 \cdot 10^{+8}$ | 1.2 MHz linewidth @ 546.1nm | [6] |
| Earth spin-down | $2.6 \cdot 10^{+12}$ | $\Delta t \approx$ 2.3 ms/cy | [103] |

**C. The Butterworth-Van Dyke (BVD) circuit**
While the RLC example is useful as an introduction to circuits, a simple modification, known as the BVD circuit, has many more applications.[88],[93] The BVD circuit consists of the series RLC circuit shunted by a parallel capacitor, $C_0$. As there is no dissipation due to $C_0$ the expressions for $\Omega^{(\pm)}$ are unaltered. The BVD circuit is used to represent many single resonance phenomena.[93] There are two capacitors in the BVD circuit, so the quantity $r = C_o/C$, comes into play. While the quantity $g = 1/[1 + Q^2 \cdot (\Omega - 1/\Omega)^2]$ does not involve r, so that $Q = 1/[\Omega^{(+)} - \Omega^{(-)}]$ may be used, many of the other network functions, such as admittance magnitude, do, and care must be exercised in using alternative definitions of Q as equal to f/Δf because "3 dB bandwidth" (Δf) becomes a property jointly of Q and r.[94] The ratio 1/r appears as a lossless coupling factor in many guises, for example in piezoelectrics and in the Lyddane-Sachs-Teller (LST) relation.[93] In these cases, the "bandwidth" arises from the coupling, and not the loss, and $1/Q = 0$.

## D. Quality factors of other phenomena

Resonance spectra having loss mechanisms are never infinitely narrow (Dirac δ-functions) but have finite bandwidths. Q, as a measure of the "quality" of the resonance, has been used in applications that vary from laser and molecular resonances to electrical circuits to electromechanical quartz oscillators to the bouncing of a golf ball on a hard surface. Some examples are shown in Table I.

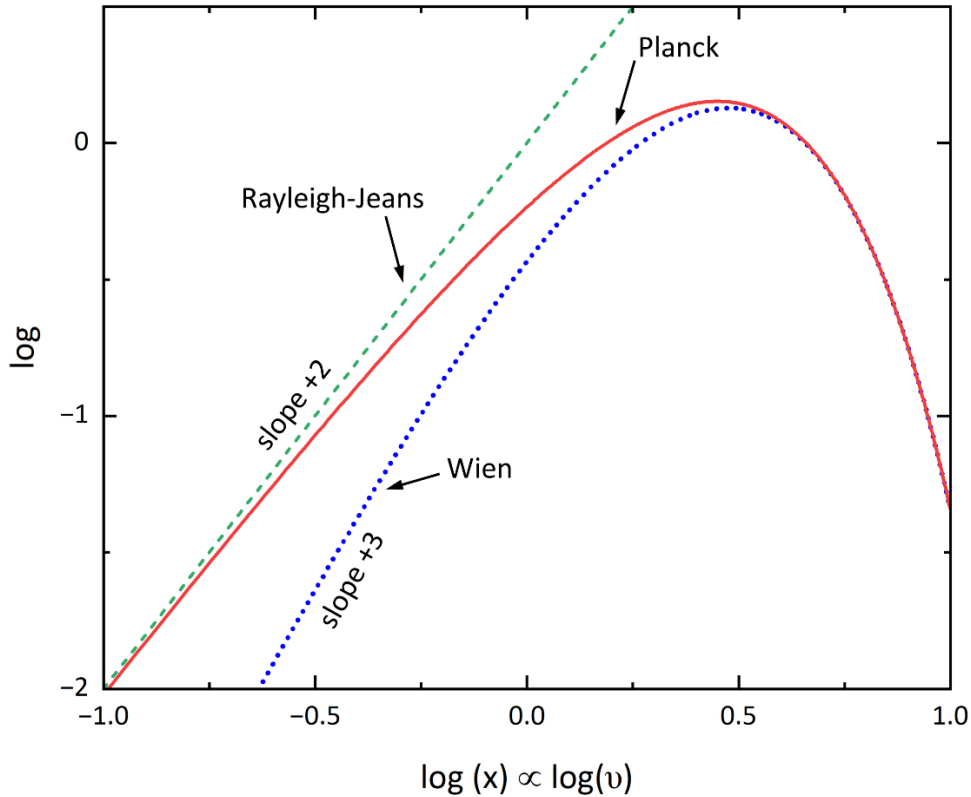

Fig. 2. Log-log plots of the Raleigh-Jeans, Wien, and Planck distribution functions. The Planck function interpolates between the other two, and yields the distribution correct at both frequency limits. The Wien function obeys Maxwell-Boltzmann statistics, whereas the Planck function obeys Bose-Einstein statistics.

## IV. BLACKBODY RADIATION – BB 101
### A. Brief synopsis of BB history

Electromagnetic (EM) radiation can be described by four attributes: wavelength, intensity, polarization, and coherence.[104],[105] In the study of blackbody radiation one seeks a relation between wavelength and intensity in the special case of radiation in steady-state equilibrium inside an isolated enclosure by an ensemble of completely incoherent harmonic oscillators, with random polarizations, at a uniform absolute temperature, T.[9],[10],[11],[13],[17],[106] Attempts at an explanation of the form of the intensity versus wavelength curve, from classical modal equipartition and thermodynamic arguments led to puzzling and contradictory results. The Rayleigh-Jeans (R-J) equipartition result [22]-[25] gave good agreement with experiment at long wavelengths, but predicted an unbounded result at short wavelengths; the famous "ultraviolet catastrophe."[26] At the other extreme, Wien's thermodynamic result [16]-[21] was in agreement with experiment at short wavelengths, but failed at longer wavelengths in what might be called the "infrared shortfall." By interpolating between these limiting

cases,[F7] thereby requiring the discretization of energy[27], Planck [28]-[31] ultimately reached the correct expression, which reduced to the earlier expressions in their correct limits, as well as to the known $T^4$ variation of total radiated power with absolute temperature, the earlier Stefan-Boltzmann law [12]-[15]. Figure 2 portrays the situation. The one pre-Planckian experimental truth that survived the quantum revolution was Wien's Displacement law: The spectral peak occurs at a wavelength ($\lambda_p$) inversely proportional to T.[8],[18]; the Stefan-Boltzmann law is a special case of the Displacement law. A much richer account of BB history is given by Richtmyer.[5]

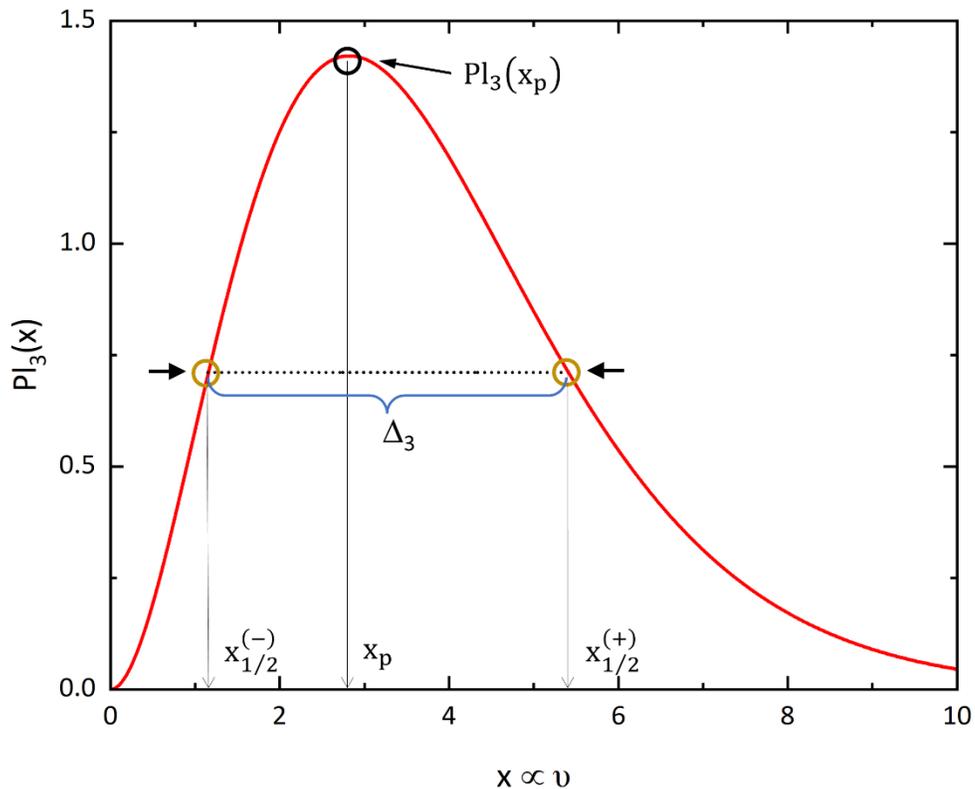

Fig. 3. Plot of the normalized quantity $Pl_3(X) = X^3/(e^X - 1)$ vs $X = h\nu/k_BT$, with ½-power points $X_{1/2}^{(\pm)}$ and ½-power bandwidth $\Delta_3$ shown. $Pl_3(X) = S_\nu/(T^3 \cdot K_1)$, where $K_1 \propto k_B^3/(c^2 \cdot h^2)$; units of $K_1$ are $[J/(m^2 \cdot K^3)]$. $S_\nu$ is a spectral radiance function.

**B. The spectral form of BB radiation**
According to the quantum view, the "shape" of any spectrum consists of a steady-state average of an innumerable number of discrete events; this being consonant with the traditional description of BB radiation as a continuous function of frequency ($\nu$) or of wavelength ($\lambda$) because of the smallness of Planck's constant. The exact shape of the blackbody spectrum depends not only on the absolute temperature, but on the dispersion (bookkeeping) rule adopted: wavelength, "intensity (or other related quantity) per unit $\lambda$ increment," or frequency, "intensity (or other related quantity) per unit $\nu$ increment" parameterization. In either case, we give the result in scaled form, such that the maxima are independent of temperature.

Any portrayal of BB spectra involves Boltzmann's constant $k_B$, Planck's constant, h, lightspeed, c, and numerics, and the admixture of these depends on the quantity described. The glossary of names for various quantities associated with BB radiation is oversized; it includes, *inter alia*: spectral radiance, emittance, excitance; monochromatic specific intensity; radiant intensity; spectral energy density; spectral radiance per unit frequency, per unit wavelength; brightness; irradiance; power intensity; etc., etc. The MKS units attached to these terms may agree or not, and one additionally finds in the literature differences in numerical factors assigned to terms bearing the same name. As we use normalized forms these terms are irrelevant for us.

Stripped to essentials, the BB spectra are given by $F_M(X) = X^M/(e^X + n)$, where X, the independent variable, equals $(h\nu/kT) = (hc/\lambda kT) = 1/Y$. M names the parametrization, with M = 3 or 5 ($\nu$ or $\lambda$ bookkeeping, respectively); the two M values differ because $\lambda \cdot \nu = c$, so the increments are related by $d\lambda = -c \cdot d\nu/\nu^2$. Coefficient n names the statistics, with n = – 1 for Planck's exact result (Bose-Einstein statistics), and n = 0 for Wien's approximation (Maxwell-Boltzmann statistics). Extensions to other M and n values are mentioned briefly in Sec. V. The peak ($X_p$) of $F_M(X)$ is found from the root of the equation $X \cdot e^X/(e^X + n) = M$, or the alternative form $n \cdot e^{-X} = [(X/M) - 1]$.[9] At the peak, $F_M(X_p) = X_p^M/(e^{X_p} + n)$. For the Planckian cases, n = – 1, and we set $F_M(X) = PI_3(\nu) = X^3/(e^X - 1)$ and $PI_5(\lambda) = X^5/(e^X - 1)$. These are graphed in Fig. 3 and Fig. 4, respectively. In each of these scaled forms, the peak is invariant with temperature.

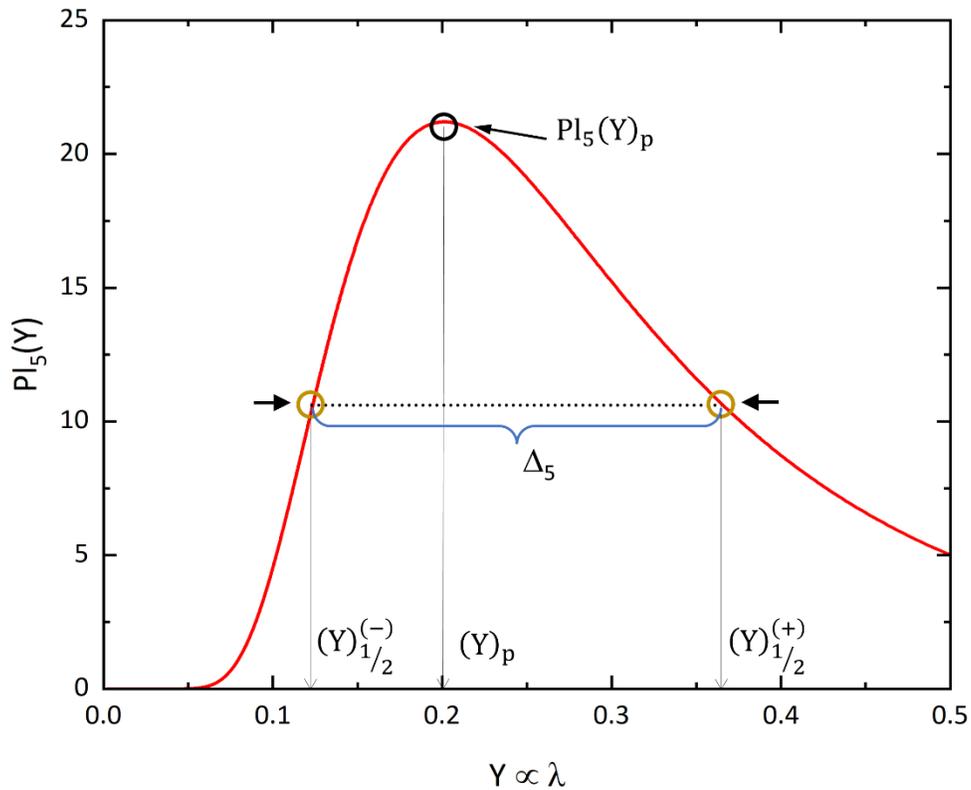

Fig. 4. Plot of the normalized quantity $Pl_5(X) = X^5/(e^X - 1)$ vs $Y = 1/X = \lambda k_B T/hc$ with ½-power points $Y_{1/2}^{(\pm)}$ and ½-power bandwidth $\Delta_5$ shown. $Pl_5(X) = S_\lambda/(T^5 \cdot K_2)$, where $K_2 \propto k_B^5/(c^3 \cdot h^4)$; units of $K_2$ are [J/(m$^3 \cdot$s$\cdot$K$^5$)]. $S_\lambda$ is a spectral radiance function.

### C. Functional behavior of the BB spectra at frequency and wavelength limits

Whereas there is an "ultraviolet catastrophe" for the R-J theory, there is no corresponding "infrared catastrophe" for the Wien theory; in the Wien theory the intensity correctly approaches zero in both high and low frequency limits. However, in the Wien theory, the functional variation goes as $\nu^3$ ($\lambda^5$) instead of the correct Planckian $\nu^2$ ($\lambda^4$) behavior. This is the "IR deficit"; the intensity approaches zero too quickly as $\nu$ diminishes to zero. Table II. shows the various limiting values.

Table II. Functional variation of the Rayleigh-Jeans, Wien, and Planck BB theories at low and high frequency limits for frequency ($\nu$) and wavelength ($\lambda$) dispersion rules.

| | $X \rightarrow 0$ | | | $Y = 1/X \rightarrow 0$ | |
|---|---|---|---|---|---|
| | Rayleigh-Jeans, Planck | Wien | | Rayleigh-Jeans | Wien, Planck |
| $\nu$ | $X^2$ | $X^3$ (IR deficit) | $\nu$ | $X^2$ (UV catastrophe) | $X^3 \cdot e^{-X}$ |
| $\lambda$ | $X^4$ | $X^5$ (IR deficit) | $\lambda$ | $X^4$ (UV catastrophe) | $X^5 \cdot e^{-X}$ |

### D. The Q concept applied to BB spectra

Applying the material discussed in Sec. III to the BB spectra in Fig. 3 and Fig. 4 permits the associated Q values to be found. The effective "Q" of a spectrum, by analogy with ordinary resonance curves such as those in Fig. 1. (b), is taken as $Q = X_p/\Delta$, where $\Delta$ is the half-maximum width ("3-dB points"). The half-maximum points are determined from the two roots $(X^{(\pm)})$ of the equation $\left(\frac{1}{2}X_p^M\right)/[e^{X_p} - 1] = (X^M)/[e^X - 1]$. The relevant quantities are given, for M = 3, in Table III, and, for M = 5, in Table IV. In both cases Q is less than 1, and interestingly, the results are <u>independent of temperature</u>.[F8] [109] One could define blackbody Q in other ways, e.g., $\nu_p/\Delta\nu$ or $\lambda_p/\Delta\lambda$ using either M = 3 or 5 values of X, etc. For all such permutations, the effective Q < 1. Given, for example in Table IV. are $Q_5(\lambda) = \lambda_p/\Delta\lambda \approx 0.831$ and $Q_5(\nu) = \nu_p/\Delta\nu \approx 0.926$.

Table III. Pertinent locations on the Planckian distribution function ($Pl_3$) with frequency parameterization (M = 3).

| Location | X | $Pl_3(X)$ |
|---|---|---|
| Lower ½-power, $X_{1/2}^{(-)}$ | $X_{1/2}^{(-)}$ = 1.1575 | 0.7107 |
| Peak | $X_p$ = 2.8214 | 1.4214 |
| 50%-area divisor | $X_{50\%}$ = 3.5030 | 1.3343 |
| Upper ½-power, $X_{1/2}^{(+)}$ | $X_{1/2}^{(+)}$ = 5.4116 | 0.7107 |
| $\Delta_3 = [X_{1/2}^{(+)} - X_{1/2}^{(-)}] \approx 4.2541$ | | $Q_3 \approx 0.663$ |

Table IV. Pertinent locations on the Planckian distribution function ($Pl_5$) with wavelength parameterization (M = 5).

| Location | Y | $Pl_5(Y)$ | X |
|---|---|---|---|
| Lower ½-power | $Y_{1/2}^{(-)}$ = 0.1235 | 10.6007 | 8.0966 |
| 50%-area divisor | $Y_{50\%}$ = 0.1779 | 20.3908 | 5.6218 |
| Peak | $Y_p$ = 0.2014 | 21.2014 | 4.9651 |
| Upper ½-power | $Y_{1/2}^{(+)}$ = 0.3660 | 10.6007 | 2.7326 |
| $\Delta_5 = [Y_{1/2}^{(+)} - Y_{1/2}^{(-)}] \approx 0.2424$ | $Q_5(\lambda) \approx 0.831$ | $Q_5(\nu) \approx 0.926$ | |

## V. COMPARISON OF STATISTICS AND M VALUES

Heald[9] considers other cases of integer dispersion rules. In addition to the rules M = 3 ($\nu$ rule) and M = 5 ($\lambda$ rule), that we have discussed, he mentions M = 2 ($\nu^2$ rule), and M = 4 (logarithmic rule = intensity per percentage bandwidth, with peak at $X_p \approx 3.9207$); of particular note is his suggestion that the median (50% of energy) rule be considered. The 50% divisor points are given in Table III. and Table IV. Another criterion of interest is the fraction of area (energy) confined between the half-power points to the total area. Examples are shown in Table V., along with the associated Q values. If M is considered as a continuous variable, then as M increases, $X_p$ approaches M from below; (for M = 4, it is within 2%); as M approaches 1 from above, $X_p$ approaches 0, and the peak value approaches 1 from below. For M $\geq$ 2, the fraction of total area between the half-power points is greater than 75%.

Table V. Fractions of areas between ½-power points to total area, and Q values associated with various line shapes. The Gaussian and Lorentzian functions describe, e.g., specific types of spectral line broadening in gases and plasmas. Doppler broadening, due to thermal motion, is represented by the Gaussian function. The Lorentzian function is used in connection with natural broadening (finite radiative lifetimes), and collisional / pressure broadening (finite lifetimes due to collisions).[F9] The first three entries are symmetrical in X; the others are not. The Gaussian ratio equals erf[ln(2)].

| Spectral line shape, normalized | | Ratio (%) | Q | Comments |
|---|---|---|---|---|
| Gaussian | $\exp(-\ln(2)\cdot x^2)$ | $\approx$ 76.10 | ½ | ½-power points at X = $\pm$ 1 |
| Lorentzian | $1/(1 + X^2)$ | 50 | ½ | ½-power points at X = $\pm$ 1 |
| RLC/BVD | $1/[1 + Q^2\cdot(X - 1/X)^2]$ | 50 | Q | variable parameter Q |
| Bose-Einstein (Planck) | $X^3/(e^X - 1)$ | $\approx$ 75.36 | 0.663 | $\nu$ dispersion |
| Bose-Einstein (Planck) | $X^5/(e^X - 1)$ | $\approx$ 75.36 | 0.831 | $\lambda$ dispersion |
| Maxwell-Boltzmann (Wien) | $X^3/(e^X - 0)$ | $\approx$ 74.81 | 0.726 | $\nu$ dispersion |
| Fermi-Dirac | $X^3/(e^X + 1)$ | $\approx$ 74.46 | 0.768 | $\nu$ dispersion |

Table VI. Q values computed by the ½-power method for three distributions, assuming frequency parameterization (M = 3). The "resonance" curve is $X^3/(e^x + n)$ versus X. For X $\rightarrow$ 0, the R-J and the asymptotic Planck slopes equal +2 on the logarithmic graph of Fig. 2., while the Wien asymptotic slope is +3. In this limit, any graph with n > 0 has an asymptotic slope of +3, but falls below the Wien curve; for example, the F-D curve has an ordinate [3·log(X) – log(2)].[111]

| Statistics | $X^3/(e^X + n)$ | $X_{1/2}^{(-)}$ | $X_p$ | $X_{1/2}^{(+)}$ | Q |
|---|---|---|---|---|---|
| Bose-Einstein | n = – 1 | 1.157465 | 2.821439 | 5.411575 | 0.6632 |
| Maxwell-Boltzmann | n = 0 | 1.394137 | 3. | 5.525350 | 0.7262 |
| Fermi-Dirac | n = + 1 | 1.536495 | 3.131020 | 5.616138 | 0.7675 |

## VI. CONCLUSION

Ascribing the EE resonance parameter "quality factor" to the graphs arising from the physics concept of BB radiation might seem more than a bit incongruous, but in the spirit of paying tribute to Kirchhoff's bicentennial, we have done just that! Apart from its obvious didactic value, the exercise reveals that the equivalent "Q" of this totally incoherent photon fluid is lower than unity, (as might be anticipated). However, this Q measure unexpectedly turns out to be independent of both the temperature, and of the dispersion rule adopted.

## AUTHOR DECLARATIONS

Conflict of Interest: The authors have no conflicts to disclose.

**a)ORCID:** https://orcid.org/0000-0003-4501-6885

**b)ORCID:** https://orcid.org/0000-0001-5910-3504

===
**Footnotes**

[F1] On pages 513 and 514 we find the enunciation of "Student" Kirchhoff's laws: "In order to be able to derive the given proportion in a convenient way, I will first prove the following theorem:
   If galvanic currents flow through a system of wires that are connected in any way, then:
   1) if the wires 1, 2, .. $\mu$ meet in one point, $I_1 + I_2 + .. + I_\mu = 0$, where $I_1, I_2,$ .. denote the intensities of the currents flowing through those wires, all counted as positive towards the point of contact;
   2) if the wires 1, 2, .. $\nu$ form a closed figure, $I_1 \cdot \omega_1 + I_2 \cdot \omega_2 + .. + I_\nu \cdot \omega_\nu =$ the sum of all electromotive forces located on the path: 1, 2, .. $\nu$ ; where $\omega_1, \omega_2, ..$ the resistances of the wires, $I_1, I_2, ..$ denote the intensities of the currents that flow through them, all counted as positive in _one_ direction."

[F2] Kirchhoff was but one of many gifted "students" who contributed to a remarkable flowering of mathematics and physics in Königsberg in the 19th century. See, e.g., Arthur Ballato, "Piezoelectricity: History and new thrusts," IEEE International Ultrasonics Symposium Proceedings (1996) 575-583. In addition to the personages noted in this paper, one might mention a few other connections. Kirchhoff was the doctoral advisor of Roland von Eötvös (PhD, 1870) and Max Noether (Dr phil., 1868). Max Noether was the father of Emmy Noether (PhD, Universität Erlangen-Nürenberg, 1907) and Fritz Noether (Dr phil., Universität München, 1909). Emmy discovered deep connections between symmetries in nature and conservation laws. See E. Noether, "Invariante Variationsprobleme," Nachrichten von der Gesellschaft der Wissenschaften zu Göttingen. Mathematisch-Physikalische Klasse. (1918) 235–257. Emmy emigrated to the US and Fritz emigrated to the USSR; he was shot by Stalin's NKVD in 1941.

[F3] Discovery of the cosmic microwave background spectrum [Robert H. Dicke, Phillip James Edwin Peebles, Peter G. Roll, and David T. Wilkinson, "Cosmic black-body radiation," Astrophysical Journal **142**(1) (1965) 414-418; Arno Allan Penzias and Robert Woodrow Wilson, "A measurement of excess antenna temperature at 4040 Mc/s," Astrophysical Journal **142**(1) (1965) 419-421] led to finding the purest example of a blackbody in nature, revealed in the results of the COBE experiment. See, e.g., John C. Mather, et al., "Measurement of the cosmic microwave background spectrum by the COBE FIRAS instrument," Astrophysical Journal **420**(2) (1994) 439-444; Dale J. Fixsen, et al., "The cosmic microwave background spectrum from the full COBE1 FIRAS data set," Astrophysical Journal **473**(2) (1996) 576-587. Additional blackbody spectra arise from neutron star collisions, see, e.g., Albert Sneppen, "On the blackbody spectrum of kilonovae," Astrophysical Journal **955**(1) (2023) 44. Blackbody radiation is truly ubiquitous; what could be more ubiquitous than the universe?

[F4] The input-output relations, for lumped, linear, finite, passive, and bilateral elements are characterized by quotients called "immittances," i.e., either impedances of admittances, defined as the ratio of an effect to its cause. For example, Ohm's law, (a zeroth-order linear DE), states that the current (i, the effect) through a resistor is proportional to the voltage (e, the cause) applied across it: $i/e = g$, the admittance.

[F5] In the sinusoidal steady-state, transients, (complementary DE solutions), are absent because of the inevitable presence of loss.

[F6] For resonances in piezoelectric materials, caution is advised, as there is an interplay between loss and piezocoupling in determining resonance widths, and thus Q. See Arthur Ballato, "Interpreting piezoceramic impedance measurements," in *Dielectric Materials and Devices*, edited by K. M. Nair, et al., (Westerville, Ohio, American Ceramic Society, 2002) 369-410.

[F7] With S the entropy of an irradiated linear, monochromatic, vibrating resonator, and U the corresponding vibrational energy, the Wien displacement law (S = f(U/ν)) implies that $\partial^2 S/\partial U^2 \propto U^{-1}$, while, for the classical harmonic oscillator, $\partial^2 S/\partial U^2 \propto U^{-2}$. Planck, the master thermodynamicist, interpolated between these by setting $\partial^2 S/\partial U^2 = \alpha/[U(U + \beta)]$, the simplest form yielding agreement with experiment, but requiring introduction of "h", a new constant of nature. See [30],[33],[34].

[F8] The temperature independence holds for the "N-dB" points, not just for the "3 dB" points; moreover, it is also independent of the M value. For the ½-power (aka 3 dB) points, $(½·X_p^M)/[\exp(X_p) – 1]$ is a constant for a given value of M. The factor ½, designating the half-power-points, could be any other value, e.g., the $1/3^{rd}$ ($\approx$ 4.77 dB) or $1/100^{th}$ (20 dB) power-points, etc. The roots become spread further apart as the dB value increases, and the definition of $Q = X_p/\Delta$ must be adjusted accordingly.

[F9] These functions may be combined, on the assumption that both mechanisms are independent, yielding the Voigt profile as their convolution. The results must be obtained by numerical integration. The 50%-area point depends, moreover, on the admixture of the distributions. See Desmond Walter Posener, "The shape of spectral lines: Tables of the Voigt profile," Australian Journal of Physics **12**(2) (1959) 184-196; Vivek Bakshi and Robert J. Kearney, "New tables of the Voigt function," Journal of Quantitative Spectroscopy and Radiative Transfer **42**(2) (1989) 111-115.

[F10] The index n may also be a rational fraction; this is a feature of quasiparticles in 2-D structures; See Jon Magne Leinaas and Jan Myrheim, "On the theory of identical particles," Il Nuovo Cimento **37B**(1) (1977) 1-23; James Nakamura, Shuang Liang, Geoffrey C. Gardner, and Michael J. Manfra, "Fabry-Pérot interferometry at the ν = 2/5 fractional quantum Hall state," Physical Review X **13**(4) (2023) 041012.